\documentclass{cupconf}
\usepackage{epsf}

  \checkfont{eurm10}
  \iffontfound
    \IfFileExists{upmath.sty}
      {\typeout{^^JFound AMS Euler Roman fonts on the system,
                   using the 'upmath' package.^^J}%
       \usepackage{upmath}}
      {\typeout{^^JFound AMS Euler Roman fonts on the system, but you
                   dont seem to have the}%
       \typeout{'upmath' package installed. cupconf.cls can take advantage
                 of these fonts,^^Jif you use 'upmath' package.^^J}%
      }
  \else
  \fi

  \checkfont{msam10}
  \iffontfound
    \IfFileExists{amssymb.sty}
      {\typeout{^^JFound AMS Symbol fonts on the system, using the
                'amssymb' package.^^J}%
       \usepackage{amssymb}%

      }{}
  \fi

  \IfFileExists{amsbsy.sty}
    {\typeout{^^JFound the 'amsbsy' package on the system, using it.^^J}%
     \usepackage{amsbsy}}
    {}

\newsavebox{\astrutbox}
\sbox{\astrutbox}{\rule[-5pt]{0pt}{20pt}}

\newcommand\farcs{\mbox{$^{\prime\prime}\!\!.$}}
\newcommand\arcsec{\mbox{$^{\prime\prime}$}}

\newcommand\ub{\mbox{$U\!-\!B$}}
\newcommand\bv{\mbox{$B\!-\!V$}}
\newcommand\vk{\mbox{$V\!-\!K$}}
\newcommand\msun{\mbox{$M_{\odot}$}}

\title{Star Formation in Clusters}

\author[S. S. Larsen]%
{S\ls {\O}\ls R\ls E\ls N\ns S.\ns L\ls A\ls R\ls S\ls E\ls N%
}

\affiliation{ESO / ST-ECF, Karl-Schwarzschild Strasse 2,
D-85748 Garching bei M{\"u}nchen, Germany}

\pubyear{2004}
\volume{xxx}
\pagerange{xxx}
\date{}
\begin{document}

\maketitle

\begin{abstract}
  The Hubble Space Telescope is very well tailored for observations of 
extragalactic star clusters. One obvious reason is HST's ability to 
recognize clusters as extended objects and measure sizes out to distances 
of several Mpc. Equally important is the wavelength range offered by the
instruments on board HST, in particular the blue and near-UV coverage 
which is essential for age-dating young clusters. 
HST observations have helped establish the 
ubiquity of young massive clusters (YMCs) in a wide variety of star-forming
environments, ranging from dwarf galaxies and spiral disks to nuclear
starbursts and mergers. These YMCs have masses and structural properties
similar to those of old globular clusters in the Milky Way and
elsewhere, and the two may be closely related. Several lines
of evidence suggest that a large fraction of all stars are born in clusters,
but most clusters disrupt rapidly and the stars disperse to become part
of the field population.  In most cases 
studied to date the luminosity functions of young cluster systems are 
well fit by power-laws $dN(L)/dL \propto L^{-\alpha}$ with $\alpha\approx2$,
and the luminosity of the brightest cluster can (with few exceptions) be 
predicted from simple sampling statistics. \emph{Mass} functions have only 
been constrained in a few cases, but appear to be well approximated by similar 
power-laws.  The absence of any characteristic mass scale for cluster 
formation suggests that star clusters of all masses form by the same basic 
process, without any need to invoke special mechanisms for the formation of
``massive'' clusters. It is possible, however, that special conditions
\emph{can} lead to the formation of a few YMCs in some dwarfs where the
mass function is discontinuous. Further studies of mass functions for
star clusters of different ages may help test the theoretical prediction 
that the power-law mass distribution observed in young cluster systems 
can evolve towards the approximately log-normal distribution seen in 
old globular cluster systems.
\end{abstract}

\firstsection 
\section{Introduction}

  The wide range of topics covered in this volume --- \emph{from
planets to cosmology} --- bear testimony to the fact that the
capabilities offered by Hubble remain unique for many purposes, in 
spite of much recent progress in competing technologies such as ground-based 
adaptive optics.  For studies of extragalactic stellar populations, the 
combination of relatively wide-field imaging at diffraction 
limited resolution even in the optical and near-UV offered by HST is 
invaluable, and will remain unsurpassed for the foreseeable 
future. 
HST has contributed much to our understanding of star formation 
in clusters, both in a Galactic and extragalactic context. Rather than
attempting to cover everything, I will concentrate mainly
on star clusters beyond the Local Group, partly because that is
what I am most familiar with, and partly because Local Group
galaxies are covered elsewhere in this volume (Grebel).  Much of the
work on young clusters done in the first decade of HST's lifetime has been 
reviewed by \cite{whit03}, and although some overlap is unavoidable 
the main aim of this review is to discuss more recent results and try 
to look ahead.  
  
  HST has had a tremendous impact on the field of extragalactic star 
clusters almost since the day it was launched. Some major ground-breaking 
discoveries were, in fact, made even prior to the 1993 repair mission.
Young star clusters had been identified in a few galaxies other than
the Milky Way prior to HST, including most Local Group
members and a few galaxies slightly beyond the Local Group. But with
HST it became possible to undertake systematic surveys in larger
samples of galaxies and better characterize the properties of young
clusters in different environments. Ground-based capabilities have also
evolved during the lifetime of HST, of course. Larger-format CCD detectors 
with excellent blue and UV quantum efficiency have become available, a
better understanding of ``dome'' seeing has led to improved image
quality, and 8--10 m ground-based telescopes have made it possible to 
obtain high-quality spectra of faint objects detected in HST images.

\section{Why star clusters?}

  While star clusters have been the subject of substantial interest 
for many years, it may be worth recalling some of the main motivations for 
studying them. 

  First, there are a number of problems which make clusters 
interesting in their own right. These involve both their formation, 
subsequent dynamical 
evolution and ultimate fate.  At first glance, clusters appear 
deceptively simple: they are aggregations of a few hundred to about a 
million individual stars, generally constituting a gravitationally bound 
system (although the latter may not be true for some of the youngest 
systems). Yet, constructing realistic models of their structure and 
dynamical evolution has proven to be a major challenge, 
and it is only now becoming possible to carry out reasonably realistic 
N-body simulations including the effects of stellar evolution, external 
gravitational fields, and the rapidly varying gravitational potential 
in the early phases of cluster evolution during which gas is expelled 
from the system (\cite[Joshi et al.\ 2000; Giersz 2001; 
Kroupa \& Boily 2002]{jos00,gie01,kb02}). The models must be tested
observationally, and HST data currently represent the only way to 
reliably measure structural parameters for extragalactic star clusters.

  Second, there is growing evidence that a significant
fraction of all stars form within clusters, although only a small
fraction of these stars eventually end up in \emph{bound} clusters
(\cite[Lada \& Lada 2003]{ll03}; \cite[Fall 2004]{fall04}). Therefore,
the problem of understanding \emph{star} formation is intimately linked to
that of understanding \emph{cluster} formation, and a theory of one
cannot be complete without the other. It is
of interest to investigate how the properties of star clusters might
depend on environment, as this might provide important clues to any
differences in the star formation process itself.  In particular, HST 
has made important contributions towards establishing the presence of 
``young massive clusters'' (YMCs\footnote{
Also known as ``super star clusters'', ``populous clusters'' or
``young globular clusters''.}) in a variety of 
environments, which appear very similar to young versions of the old 
\emph{globular} clusters which are ubiquitous around all major galaxies. 
Globular cluster formation was once thought to be uniquely related to the 
physics of the early Universe 
(\cite[e.g.\ Peebles \& Dicke 1968; Fall \& Rees 1985]{pd68,fr85}) but it 
now seems to be an ongoing process which can be observed even at the 
present epoch. 

  Third, star clusters are potentially very useful as tracers of the
stellar populations in their host galaxies. Clusters can be 
identified and studied at much greater distances than individual stars.
In most cases, they are composed of stars which, to a very good 
approximation, formed at the same time and have the same metallicity. This 
is in contrast to the integrated light from the galaxies, which may 
originate from an unknown mix of stellar populations with different ages 
and metallicities.  Although the effects of stellar evolution alone cause 
a cluster to fade by 5-6 magnitudes (in $V$-band) over 10 Gyrs
(\cite[Bruzual \& Charlot 2003]{bc03}), it is in principle possible
to detect clusters which have formed during the entire lifetime
of galaxies, out to distances of several Mpc. In particular, \emph{globular 
clusters} have been used extensively in attempts to constrain the 
star formation histories of early-type galaxies.

\section{HST and Extragalactic Star Clusters}
\label{sec:hst_esc}

  HST is almost ideally tailored for studies of extragalactic star
clusters. Three main reasons for this are:

\begin{itemize}
  \item Angular resolution: clusters typically have half-light radii of 
      2--4 pc (see Section~\ref{sec:env}), and can thus be recognized as 
      extended objects out to 
      distances of 10--20 Mpc with the $\sim0\farcs05$ resolution offered
      by WFPC2 or ACS. With careful modeling of the point spread function
      (PSF) this limit may be pushed even further.
  \item Field size: At 10 Mpc, the ACS $200\arcsec\times200\arcsec$ 
      field-of-view corresponds to about 10 kpc $\times$ 10 kpc, 
      making it possible to cover a significant fraction of a typical
      galaxy in a single pointing.
  \item Spectral range: For studies of young stellar populations,
      optical and near-UV spectral coverage is essential, as 
      discussed below.
\end{itemize}

There is currently no alternative to HST on the horizon
which offers a similar combination of capabilities.
The \emph{James Webb Space Telescope} (JWST), while offering vastly 
improved efficiency in the IR, will offer no significant gain in
resolution over HST, and will be limited to longer wavelengths. Ground-based
adaptive optics (AO) can provide similar, or even better resolution than
HST, but only within a small ($\sim 20\arcsec$) isoplanatic field of view.
Furthermore, AO lacks the stable PSF of HST
which is critical for many purposes (e.g.\ when measuring structural
parameters for star clusters at the limit of the resolution), and
is in any case limited to the IR (at least for now). The GALEX mission 
offers wide-field UV imaging, but with a spatial resolution that is 
inferior by far to that of HST (about 5$\arcsec$). 

\begin{figure}
\centerline{
  \epsfxsize=9cm
  \epsfbox{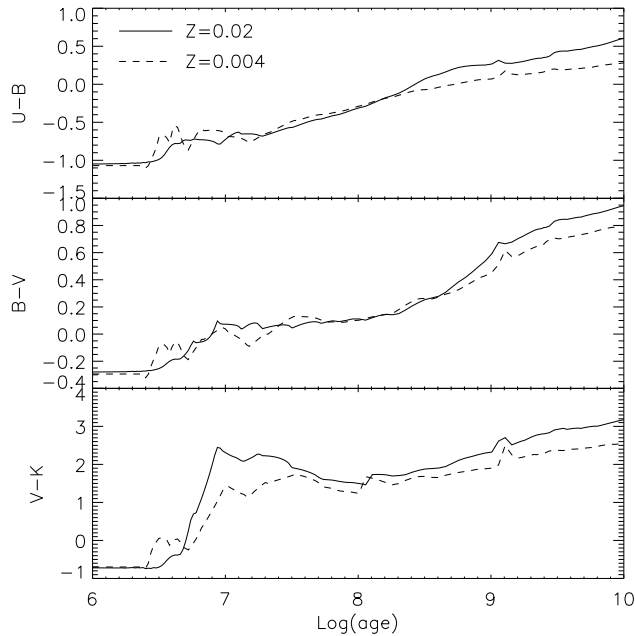}
}
\caption{\label{fig:col_age}Evolution of broad-band colors as
a function of age and metallicity according to Bruzual \& Charlot (2003)
simple stellar population models.
}
\end{figure}

The need for optical and near-UV imaging in particular deserves some 
additional comments. Figure~\ref{fig:col_age} shows simple stellar
population (SSP) model calculations (\cite[Bruzual \& Charlot 2003]{bc03}) 
for the evolution of the \ub, \bv\ and \vk\ broad-band
colors of a single-burst stellar population.  The models are shown for 
metallicities $Z=0.02$ (Solar) and $Z=0.004$ between ages of $10^6$ 
years and $10^{10}$ years.  As seen from the figure, the \ub\ color is an 
excellent age indicator in the range from $10^7$ to a few times $10^8$ years, 
increasing by more than 0.5 mag and with little metallicity dependence over
this age range.
The \bv\ color, in contrast, remains nearly constant over the same
age range, and offers little leverage for age determinations. 
In practice, there are complicating problems such as dust extinction, 
which in general will make it difficult to obtain accurate age estimates 
from a single color. Using a combination of two colors, such as \ub\ and 
\bv, will make it possible to constrain both age and reddening,
while at the same time being relatively insensitive to metallicity
effects. The relation between age and location of a cluster 
in the (\ub,\bv) two-color plane has been calibrated with clusters in
the Large Magellanic Cloud through the so-called `S'-sequence
(\cite[Elson \& Fall 1985]{ef85}; \cite[Girardi et al. 1995]{gir95}).
For ages younger than about $10^7$ years, line emission becomes important
(\cite[Anders \& Fritze-v.\ Alvensleben 2003]{af03}), while the
age-metallicity degeneracy (\cite[Worthey 1994]{wor94}) becomes a 
difficulty at older ages.
A more recent discussion of photometric age indicators, with emphasis
on the importance of blue and UV data, is in
\cite{and04b}.  Use of e.g.\ the \vk\ color can help put further constraints 
on the metallicity and may also help constrain the ages of stellar
populations in the range $\sim200$ Myr to $\sim500$ Myr
(\cite[Maraston et al.\ 2002]{mar02}),
although the models are more uncertain and 
depend strongly on the stellar evolutionary tracks used in 
the construction of the SSP models (\cite[Girardi 2000]{gir00}).

High-resolution, wide-field imaging in the blue and/or $UV$ will be
especially important for attempts to constrain not only the
luminosity function, but also the mass function of clusters.
For a long time, WFPC2 was the ``workhorse'' on HST, and it remains
the only wide-field imager on board HST with $U$-band imaging
capability. However,
the sensitivity of WFPC2 in the $U$-band is rather low and the 
detectors are steadily degrading. The 
\emph{Wide Field Camera 3}, with its panchromatic coverage, would 
be an ideally suited instrument for such studies.


\section{Setting the stage: early developments}
\label{sec:early}

  Even within the Local Group, it has long been known that the
traditional distinction between open and globular clusters that can
be applied fairly easily in the Milky 
Way breaks down in some other galaxies.  The classical example is the ``blue
globular'' clusters in the Large Magellanic Cloud, which are not easily 
classified as either open or globular clusters.
The most massive of these objects have masses up to 
$\sim10^5 \msun$ 
(\cite[Elson \& Fall 1985; Fischer et al.\ 1992; Richtler 1993;
 Hunter et al. 2003]{ef85,fis92,ric93,hun03})
similar to the median mass of old globular clusters 
and about an order of magnitude more massive than any young open cluster 
known in the Milky Way. Yet, these objects have young ages, and are
still being produced today by the LMC.  Similar clusters
have been found in M33 (\cite[Christian \& Schommer 1982,1988]{cs82,cs88}).

  A good indication of the status of research in extragalactic young
star clusters shortly prior to HST is provided by 
\cite[Kennicutt \& Chu (1988; hereafter KC88)]{kc88}. These authors compiled 
observations of what they refer to as ``young populous clusters'' (PCs) in 
14 galaxies for which data was available at that 
time. Half of the galaxies studied by KC88 were Local Group members (Milky 
Way, LMC, SMC, M33, M31, 
NGC~6822 and IC~1613). As noted by KC88, a severe difficulty in comparing
observations of PCs in different galaxies, made by 
different authors, is the widely variable completeness of the surveys, and 
the different definitions of such clusters. 
KC88 adopted a (somewhat arbitrary) definition of a young PC 
as an object with an estimated mass $>10^4\,\msun$ and a color $\bv<0.5$.
They noted a conspicuous deficiency of populous clusters in the Milky Way and 
M31, the two only large Sb/Sbc-type spirals in the sample, and suggested that 
this might be linked to the deficiency of giant H{\sc ii} regions in the same
two galaxies. By comparing the relative numbers of PCs and giant
H{\sc ii} regions in their sample of galaxies, KC88 concluded that PCs 
may indeed form inside such regions, but not all giant H{\sc ii} regions 
produce bound clusters. This is very much in line with recent indications
that only a small fraction of star clusters of \emph{any} mass remain
bound (\cite[Fall 2004]{fall04}).  The galaxies which did contain PCs were all 
late-type, though not \emph{all} late-type galaxies were found to contain
PCs.  A significant exception is the Local Group dwarf irregular IC~1613, 
which contains few if any star clusters at all 
(\cite[van den Bergh 1979; Hodge 1980]{van79,hod80}) in spite of some
on-going star formation. The near-absence of star clusters in IC~1613 may
be as important a clue to the nature of the cluster formation process
as the abundant cluster systems in starbursts and merger galaxies
(Section \ref{sec:env}).

  To a large extent, research in old globular clusters (GCs) remained 
detached from that of YMCs until fairly recently. It
was well-known that early-type galaxies typically have many more
GCs per unit host galaxy luminosity 
(\cite[Harris \& van den Bergh 1981]{hv81}) than spirals and irregulars, 
a fact that was
recognized as a problem for the idea that early-type galaxies
form by mergers of gas-rich spirals (\cite[van den Bergh 1982]{van82}).  
\cite{schw87} 
proposed that this problem might be solved if new GCs form \emph{during} 
the merger.  This idea was further explored by \cite{az92}, who
predicted that the resulting merger product should contain two distinct 
GC populations: one metal-poor population inherited from the progenitor 
galaxies, and a new metal-rich population formed in the merger. The two 
GC populations should be identifiable in the color distributions of the
resulting GC 
systems. Two highly influential discoveries soon followed: Bimodal color 
distributions were discovered in several GC systems around early-type galaxies 
(\cite[Zepf \& Ashman 1993; Secker et al.\ 1995; 
Whitmore et al.\ 1995]{za93,sec95,whit95}), and highly 
luminous, compact young star clusters 
were found in ongoing or recent mergers like the Antennae and NGC~7252
(\cite[Whitmore et al.\ 1993; Whitmore \& Schweizer 1995]{whit93,ws95}).
In retrospect, it had already been known for a long time that even
the metallicity distribution of the Milky Way GC system is strongly bimodal
(\cite[Zinn 1985]{zinn85}). The mean metallicities of the two modes in
the Milky Way are 
in fact quite similar to those seen in early-type galaxies.
The Milky Way is unlikely to be the result of a major merger, and
there are also other indications that not all properties of GC systems 
in early-type galaxies can be explained by a naive application of the
merger model. Alternative scenarios have later been put forward to explain 
the presence of multiple GC populations 
(\cite[e.g.\ Forbes et al.\ 1997; C{\^o}t{\'e} et al.\ 1998]{for97,cot98}),
but it is beyond the scope of this paper to discuss any of these in
detail. Comprehensive discussions can be found e.g.\ in \cite{whar01}
and \cite{kp00}.
Nevertheless, the discovery of young globular cluster-like objects in 
ongoing mergers was a tantalizing hint that it might be possible to study 
the process of globular cluster formation close-up at the present epoch, 
and not just from the fossil record.


\section{Extragalactic Star Clusters in Different Environments}
\label{sec:env}

Tables~\ref{tab:nmsb}--\ref{tab:sp} are an attempt to collect a reasonably 
complete list of galaxies where YMCs have been identified (until
~$\sim$May 2004), along with 
some pertinent references. For each galaxy, the main facilities used for
the observations are listed, although in many cases it is impossible to 
give a comprehensive listing. Standard abbreviations (ACS, FOC, GHRS, STIS, 
NICMOS, WFPC, WFPC2) are used for HST instruments. Other abbreviations
are WIYN (Wisconsin Indiana Yale NOAO 3.5 m), 
UKIRT (United Kingdom Infra-Red Telescope), NTT 
(ESO 3.5 m New Technology Telescope), CFHT 
(3.6 m Canada-France Hawaii Telescope), DK154 (Danish 1.54 m at ESO, La Silla)
and NOT (2.56 m Nordic Optical Telescope).  The level of detail provided in 
different studies varies enormously -- in some cases, identifications of YMCs 
are only a byproduct of more general investigations of galaxy properties
(\cite[e.g.\ Meurer et al.\ 1995; Scoville et al.\ 2000]{meu95,sco00})
while other studies are dedicated analyses of cluster systems in
individual galaxies. Galaxies marked with an asterisk ($\star$) are
used later (Section~\ref{sec:size_sample}) when discussing luminosity functions.
In the following I briefly discuss a few illustrative
cases from each table and then move on to discuss more general
properties of young cluster systems.
  
\subsection{Starburst galaxies}
\label{sec:nmsb}

\begin{table}
\begin{center}
{\indexsize
\begin{tabular}{lp{3cm}p{7cm}}
  Galaxy    & Instrument & References \\[3pt]
  NGC 3125  & STIS    & \cite{chan04} \\
  NGC 3310$^\star$  & FOC, WFPC2, NICMOS & \cite{meu95,deg03a} \\
  NGC 3991$^\star$  & FOC     & \cite{meu95} \\
  NGC 4670$^\star$  & FOC     & \cite{meu95} \\
  NGC 5253$^\star$  & STIS, WFPC2, FOC & \cite{ma01,tre01,har04,tb04,vs04,meu95} \\
  NGC 6745   & WFPC2 & \cite{deg03a} \\
  NGC 7469   & NICMOS & \cite{sco00} \\
  NGC 7673$^\star$   & WIYN, WFPC2 & \cite{hg99,hom02} \\
  IC 883     & NICMOS & \cite{sco00} \\
  M 82       & Pal 200-inch,WFPC, WFPC2 & \cite{van71,oco95,deg03b} \\
  TOL1924-416$^\star$     & FOC   & \cite{meu95} \\
  Zw 049.057      & NICMOS   & \cite{sco00} \\
  VII Zw 031      & NICMOS   & \cite{sco00} \\
  IR 15250+3609   & NICMOS   & \cite{sco00} \\
  IR 17208-0014   & NICMOS   & \cite{sco00} \\
  NGC 1614        & NICMOS, UKIRT & \cite{ah01a,kot01} \\
  NGC 7714        & UKIRT JHK   & \cite{kot01} \\
\end{tabular}
}
\caption{Observations of young star clusters in starburst galaxies.}
\label{tab:nmsb}
\end{center}
\end{table}

\begin{table}
\begin{center}
{\indexsize
\begin{tabular}{lp{4cm}p{7cm}}
  Galaxy    & Instrument & References \\[3pt]
  NGC 253$^\star$   & WFPC2, ROSAT & \cite{wat96,for00} \\
  NGC 1808  & NTT K-band & \cite{tac96} \\
  NGC 4303  & WFPC2, NICMOS, STIS & \cite{cw00,col02} \\
  NGC 5236  & WFPC2 & \cite{har01} \\
  NGC 6240  & WFPC2, NICMOS & \cite{pas04,sco00} \\
  NGC 1079  & FOC   & \cite{mao96} \\
  NGC 1326$^\star$  & WFPC2 & \cite{but00} \\
  NGC 1433  & FOC   & \cite{mao96} \\
  NGC 1512$^\star$  & FOC, WFPC2, NICMOS & \cite{mao96,mao01a} \\
  NGC 1097$^\star$  & WFPC2 & \cite{bar95} \\
  NGC 2903  & CFHT/AO NICMOS & \cite{ah01b} \\
  NGC 2997  & FOC, WFPC2 & \cite{mao96} \\
  NGC 3310$^\star$  & KPNO 4m, WFPC2 & \cite{elm02} \\
  NGC 4314$^\star$  & WFPC & \cite{ben93} \\
  NGC 5248$^\star$  & FOC, WFPC2, NICMOS & \cite{mao96,mao01a} \\
  NGC 6951$^\star$  & WFPC2 & \cite{bar95} \\
  NGC 7552$^\star$  & FOC & \cite{meu95} \\
\end{tabular}
}
\caption{Observations of young star clusters in nuclear and circumnuclear 
  starbursts.}
\label{tab:nucsb}
\end{center}
\end{table}

  The richest populations of YMCs are often found in 
major mergers (Section~\ref{sec:mergers}).  However, there are also 
examples of YMCs in starbursts which are not directly associated with major 
mergers, although they may in some cases be stimulated by more benign 
interactions or accretion of companion satellites
(Table~\ref{tab:nmsb}).  In the case of NGC~7673, for example, \cite{hg99} 
argue that the morphological features of the galaxy point toward a
minor merger, while the starburst in M~82 may have been triggered
by tidal interactions with M~81.  M~82 is also noteworthy for being 
the first galaxy in which the term `super star cluster' was used. 
It was introduced by \cite{van71}, who was careful to point out that 
the nomenclature was not intended to imply that these objects are 
necessarily bound. The presence of SSCs in M~82 was confirmed by
\cite{oco95} who identified about 100 clusters in WFPC images.
An example of a starburst which
is unlikely to be triggered by an interaction is NGC~5253, which is
located about 600 kpc from its nearest neighbor, M~83 
(\cite[Harris et al.\ 2004]{har04}). 

One of the first surveys to provide a systematic census of star 
clusters in a sample of starburst galaxies was the work by 
\cite{meu95}, who observed 9 galaxies with HST's 
Faint Object Camera (FOC).  Meurer et al.\ noted that a high fraction, 
on average about 20\%, of the UV luminosity in these starbursts originated 
from clusters or compact objects, and a hint of a trend for this fraction 
to increase with the underlying UV surface brightness.  They also measured 
cluster sizes similar to those of Galactic globular clusters, and found the 
luminosity functions to be well represented by a power-law 
$dN(L)/dL \propto L^{-\alpha}$ with $\alpha\approx2$. 

  YMCs have been identified in several nuclear and circumnuclear starburst 
regions, often associated with barred spiral galaxies
(Table~\ref{tab:nucsb}). \cite{mao96} studied 
5 circumnuclear starbursts and found that as much as 30\%--50\% of the UV 
light came from compact, young star clusters with half-light radii
$<5$ pc and estimated masses up to about $10^5\,\msun$. Again they found the 
luminosity functions to be well approximated by power-laws with slope 
$\alpha\approx2$.  \cite{but00} found a much steeper slope 
($\alpha=3.7\pm0.1$) in their study of the circumnuclear starburst in 
NGC~1326, but noted that their sample might be contaminated by individual 
supergiant stars. In some cases the ring-like structure of the nuclear
starburst is not quite so evident. \cite{wat96} discovered 4 luminous 
clusters in the central starburst region of NGC~253, the brightest of 
which has $M_V=-15$, an inferred mass in excess of 
$1.5\times10^6\,\msun$ and a half-light radius of 2.5 pc. However,
these clusters may be part of a compact ring-like structure with a radius
of about 50 pc (\cite[Forbes et al.\ 2000]{for00}). Most of the clusters 
in the nuclear starburst of M83 are also located within a semicircular
annulus (\cite[Harris et al.\ 2001]{har01}), but again the ring
is more poorly defined.

\subsection{Mergers}
\label{sec:mergers}

\begin{table}
\begin{center}
{\indexsize
\begin{tabular}{lp{3cm}p{7cm}}
  Galaxy    & Instrument & References \\[3pt]
  NGC 520         & UKIRT JHK & \cite{kot01} \\
  NGC 1275$^\star$  & WFPC, WFPC2, Keck/LRIS & \cite{hol92,car98,bro98} \\
  NGC 1741$^\star$  & FOC, WFPC2, GHRS & \cite{jon99} \\
  NGC 2207$^\star$ / IC 2163  & WFPC2 & \cite{elm01} \\
  NGC 2366$^\star$  & WFPC2, STIS, CFHT H$\alpha$, JHK & \cite{dri00} \\
  NGC 2623  & NICMOS  & \cite{sco00} \\
  NGC 3256$^\star$  & WFPC2  & \cite{zep99} \\
  NGC 3395/96 & STIS & \cite{han03} \\
  NGC 3597$^\star$  & ESO 2p2, NTT, WFPC2 & \cite{lut91,hol96,car99,fh00} \\
  NGC 3690$^\star$  & FOC  & \cite{meu95} \\
  NGC 3921$^\star$  & WFPC2  & \cite{sch96} \\
  NGC 4038/39$^\star$ & WFPC, WFPC2, GHRS & \cite{ws95,whit99} \\
  NGC 6052$^\star$  & WFPC2 & \cite{hol96} \\
  NGC 6090  & NICMOS & \cite{din99,sco00} \\
  NGC 6240  & WFPC2  & \cite{pas03} \\
  NGC 7252$^\star$  & WFPC, WFPC2  & \cite{whit93,mil97} \\
  NGC 7727  & ?  & \cite{cs94} \\
  Arp 220    & NICMOS & \cite{sco98,sco00} \\
  II ZW 96  & Univ. Haw. 2p2 BRHK  & \cite{gol97} \\
  The Mice  & ACS  & \cite{deg03c} \\
  Tadpole   & ACS  & \cite{deg03c} \\
  HCG 31    & WIYN, WFPC2  & \cite{jc00} \\
  VV 114E/W & NICMOS  & \cite{sco00} \\
  UGC 5101  & NICMOS  & \cite{sco00} \\
  UGC 10214 & ACS  & \cite{tra03} \\
  IR 10565+2448W & NICMOS & \cite{sco00} \\
  IR 15206+3342  & WHT, WFPC2 & \cite{ac02} \\
  IR 22491-1808W & NICMOS& \cite{sco00} \\
  Mrk 273S     & NICMOS& \cite{sco00} \\
  Stephan's Quintet  & WFPC2 & \cite{gal01} \\
  Tidal Tails    & WFPC2 & \cite{kni03} \\
\end{tabular}
}
\caption{Observations of young star clusters in mergers.}
\label{tab:mergers}
\end{center}
\end{table}

  Many of the most spectacular YMC populations have been found in
merger galaxies. NGC~1275 was one of the first galaxies in which
HST data confirmed the existence of YMCs, although at least one
object in this galaxy was already suspected to be a massive cluster based on 
ground-based data \cite[(Shields \& Filippenko 1990)]{sf90}. With the
Planetary Camera on HST, \cite{hol92} identified about 60 cluster
candidates with absolute magnitudes up to $M_V=-16$.
Using WFPC2 data, \cite{car98} identified about 3000 
clusters, of which about 1200 have blue integrated 
colors and estimated ages between 0.1 and 1 Gyr. The young clusters
had estimated masses and sizes similar to those of old globular clusters,
although \cite{bro98} found that the Balmer line equivalent widths
measured on spectra of 5 clusters were too strong to be consistent with 
standard SSP models, unless a stellar mass function truncated at 
$2\msun-3\msun$ 
was adopted. With accurate modeling of the HST point spread function
and high dispersion spectroscopy with 8--10 m class telescopes, it might
be possible to constrain the virial masses of some of the brightest clusters, 
and thereby provide independent constraints on their stellar IMF.

  While NGC~1275 may have experienced a recent merger / accretion
event (\cite[Holtzman et al.\ 1992]{hol92}), it is hardly one of the
classical ``Toomre'' mergers 
(\cite[Toomre \& Toomre 1972]{tt72}).  One of the nearest ongoing, major
mergers is the ``Antennae'' NGC 4038/39, where HST observations have
revealed a rich population of luminous, compact young star clusters with
typical half-light radii $\sim4$ pc
(\cite[Whitmore \& Schweizer 1995; Whitmore et al.\ 1999]{ws95,whit99}).
The brightest of them reach $M_V\approx-14$ and have estimated
masses close to $10^6\msun$ (\cite[Zhang \& Fall 1999]{zf99}). 
Similar rich populations of YMCs have been found in many other mergers,
like NGC~3256 where \cite{zep99} identified about 1000 compact bright, blue 
objects on WFPC2 
images within the central 7 kpc $\times$ 7 kpc region. Again, the
young clusters contribute a very significant fraction (15\%--20\%) of the
blue light within the starburst region. \cite{zep99} estimated half-light
radii of 5--10 pc for the clusters in NGC~3256, somewhat larger than
for the Antennae, but note that 1 PC pixel corresponds to a linear scale
of 8 pc at the distance of NGC~3256, so that the clusters are only
marginally resolved. Interestingly, only a shallow trend of cluster
size versus luminosity was found, with radius $r$ scaling with
luminosity $L$ roughly as $r \propto L^{0.07}$.

  NGC~7252 is a somewhat more advanced system than NGC~3256 or the
Antennae. \cite{mil97} date the cluster system at between
650 Myr and 750 Myr. Remarkably, both photometry and dynamical
measurements yield a mass of about $8\times10^7\,\msun$ for the
most massive object (W3) (\cite[Maraston et al.\ 2004]{mar04}), making
it about an order of magnitude more massive than any old globular cluster in 
the Milky Way. With a half-light radius of $17.5\pm1.8$ pc, this
object is much larger than a normal star cluster, and may be
more closely associated with the ``Ultra Compact Dwarf Galaxies'' in 
Fornax (\cite[Hilker et al.\ 1999; Drinkwater et al.\ 2003]{hil99,dri03}).

\subsection{Dwarf / Irregular galaxies}

\begin{table}
\begin{center}
{\indexsize
\begin{tabular}{lp{3cm}p{7cm}}
  Galaxy    & Instrument & References \\[3pt]
  NGC 1140$^\star$  & WFPC  & \cite{hun94} \\
  NGC 1156$^\star$  & NOT, WFPC2   & \cite{lr99,lar02} \\
  NGC 1313$^\star$  & DK154, WFPC2 & \cite{lr99,lar02} \\
  NGC 1569$^\star$  & Pal 200-inch, WFPC, WFPC2, STIS & 
               \cite{as85,oco94,dem97,hun00,mao01b,ma01,ori01,and04a,gg03} \\
  NGC 1705$^\star$  & Las Campanas 2p5, WFPC, WFPC2, STIS & 
               \cite{mel85,oco94,ma01,bil02,vaz04} \\
  NGC 3077$^\star$  & WFPC2, CFHT JHK & \cite{har04,dav04} \\
  NGC 4194  & STIS & \cite{wei04} \\
  NGC 4214$^\star$  & WFPC2 & \cite{ma01,bil02} \\
  NGC 4449$^\star$  & WFPC2 & \cite{gel01,ma01} \\
  ESO-338-IG04 & WFPC2 & \cite{ost98} \\
  HE 2-10   & WFPC2, GHRS & \cite{cv94,jo00,be01} \\
  I Zw 18$^\star$   & FOC & \cite{meu95} \\
  UGC 7636  & KPNO 4m C,T1 & \cite{lee97} \\
  POX 186   & NTT RI & \cite{dou00} \\
  SBS 0335-052 & NTT JHK & \cite{van00} \\
\end{tabular}
}
\caption{Observations of young star clusters in dwarf and irregular 
  galaxies.}
\label{tab:dwsb}
\end{center}
\end{table}

The bright ``central condensations'' in NGC~1569 were noted
already by \cite{ma35} on plates taken with the 36 inch Crossley
reflector at Lick Observatory, though \cite{as85} were probably the first
to recognize them as likely star clusters. At a distance of only
$\sim2$ Mpc (\cite[Makarova \& Karachentsev 2003)]{mk03}, these clusters 
appear well resolved on HST images with half-light radii of about 2 pc 
(\cite[O'Connell et al.\ 1994; de Marchi et al.\ 1997]{oco94,dem97}).
One of the clusters, NGC~1569-A, is actually a double cluster, and
STIS spectroscopy has shown that one component exhibits
Wolf-Rayet features while the other component is devoid of such
features, suggesting an age difference of a few Myrs between
the two components (\cite[Maoz et al.\ 2001b]{mao01b}). 
Using high-dispersion spectroscopy from the NIRSPEC spectrograph
on the Keck II telescope, \cite{gg03} derived dynamical mass estimates
of about $0.3\times10^6 \msun$ for each of the two components of
NGC~1569-A, and $0.18\times10^6\msun$ for NGC~1569-B, again very
similar to the typical masses of old globular clusters, and
consistent with the clusters having ``normal'' stellar mass functions
(see also Section \ref{sec:eff}).

A peculiar feature of the NGC~1569 cluster population
is that the next brightest clusters after NGC~1569-A and NGC~1569-B
are more than 2 magnitudes fainter (\cite[O'Connell et al.\ 1994]{oco94}).
An even more dramatic discontinuity in the luminosity function is seen in 
NGC~1705 which has only a single bright cluster, and
in NGC~4214 there is a gap of about 1.5 mag from the
brightest 2 clusters down to number 3 (\cite[Billett et al.\ 2002]{bil02}).
Interestingly, while the clusters in NGC~1569 and NGC~1705 are
young ($\sim10^7$ years), the two clusters in NGC~4214 are both
about 250 Myrs old (\cite[Billett et al.\ 2002]{bil02}), demonstrating
that massive clusters are capable of surviving for substantial amounts of 
time at least in some dwarf galaxies. 

\subsection{Spiral galaxy disks}

\begin{table}
\begin{center}
{\indexsize
\begin{tabular}{lp{3cm}p{8cm}}
  Galaxy    & Instrument & References \\[3pt]
  M 51$^\star$      & Lick 3m UBV, WFPC2, NICMOS & \cite{lar00,bik03,bas04} \\
  M 81      & WFPC2  & \cite{chan01} \\
  M 101     & WFPC2  & \cite{bre96} \\
  NGC 2403$^\star$  & Loiano 1p5, WFPC2, NOT & \cite{bat84,dri99,lr97} \\
  NGC 2997$^\star$  & DK154, WFPC2  & \cite{lr99,lar02} \\
  NGC 3081  & WFPC2  & \cite{but04} \\
  NGC 3621$^\star$  & DK154, WFPC2 & \cite{lr99,lar02} \\
  NGC 3627$^\star$  & WFPC2  & \cite{dk02} \\
  NGC 5236$^\star$  & UITP$^1$, DK154, WFPC2  & \cite{boh90,lr99,lar02} \\
  NGC 7793$^\star$  & DK154, WFPC2  & \cite{lr99,lar02} \\
  NGC 6946$^\star$  & NOT, WFPC2  & \cite{lr99,lar02} \\
\end{tabular}
}
\caption{Observations of young star clusters in spiral galaxy disks. 
 $^1$UITP = prototype of the Ultraviolet Imaging Telescope flown on Spacelab.}
\label{tab:sp}
\end{center}
\end{table}

  Most of the YMCs discussed in the preceding sections are located
in environments that are peculiar in some way, or at least different
from what we see in the solar neighborhood. Thus, it is tempting to 
speculate that the absence of YMCs in the Milky Way indicates that
their formation somehow requires special conditions. 
There is, however, increasing evidence that YMCs can form even in the 
disks of spiral galaxies. Table~\ref{tab:sp} lists a number of nearby 
spirals in which YMCs have been identified.  A few (e.g.\ M51) are 
clearly involved in interactions, but none of them are disturbed to 
a degree where they are not clearly recognizable as spirals. The nuclear 
starburst in M~83 was already mentioned in Section~\ref{sec:nmsb}, but 
there is also a rich population of young star clusters throughout the
disk (\cite[Bohlin et al.\ 1990; Larsen \& Richtler 1999]{boh90,lr99}), 
the most massive of which have masses of several times $10^5\,\msun$.
An even more extreme cluster is in NGC~6946, with a dynamical mass estimate
of about $1.7\times10^6\,\msun$ (\cite[Larsen et al.\ 2001]{lar01}).
The disks of spiral galaxies can evidently form star clusters
with masses as high as those observed in any other environment, 
including merger galaxies like the Antennae and starbursts like M~82.

Most of the spirals in Table~\ref{tab:sp} are type Sb or later, but
one exception is NGC~3081. In this barred S0/Sa-type spiral, 
\cite{but04} detected a number of luminous young clusters in the
inner Lindblad resonance ring at 5 kpc. \cite{but04}
found rather large sizes for these clusters, with estimated half-light
radii of about 11 pc. This is much larger than the typical sizes
of Milky Way open and globular clusters and indeed of YMCs found
in most other places, and raises the question whether these objects
might be related to the ``faint fuzzy'' star clusters which are
located in an annulus of similar radius in the lenticular galaxy
NGC~1023 (\cite[Larsen \& Brodie 2000; Brodie \& Larsen 2002]{lb00,bl02}),
but have globular cluster-like ages.

\section{General properties of cluster systems}

  Just how similar are the properties of star clusters in different 
environments, and what might they tell us about the star formation process?  
Objects like NGC1569-A appear extreme compared to Milky Way open clusters
or even to young LMC
clusters: \cite{oco94} estimate that NGC1569-A has a half-light surface 
brightness over 65 times higher than the R136 cluster in the LMC, 
and 1200 times higher than the mean rich LMC cluster after allowing for 
evolutionary fading.  Do such extreme objects constitute an altogether
separate mode of star/cluster formation, or do they simply represent
a tail of a distribution, extending down to the open clusters
that we encounter locally? And are YMCs really
young analogs of the old GCs observed in the Milky Way and virtually
all other major galaxies?

\subsection{Luminosity- and mass functions}
\label{sec:lfmf}

  One of the best tools to address these questions is the cluster mass function 
(MF).  In the Milky Way and the Magellanic Clouds, the MF of young star 
clusters is well approximated by a power-law $dN(M)/dM \propto M^{-\alpha}$ 
where $\alpha\approx2$ (\cite[Elmegreen \& Efremov 1997; 
Hunter et al.\ 2003]{ee97,hun03}). This is deceptively similar to the 
\emph{luminosity} functions derived in many young cluster systems,
but it is important to recognize that luminosity 
functions are not necessarily identical, or even similar to the underlying 
MFs (unless the age distribution is a delta function).  
Unfortunately, MFs are difficult to measure directly. The 
only practical way to obtain mass estimates for large samples of clusters 
is from photometry, but because the mass-to-light ratios are strongly 
age-dependent masses cannot be estimated without reliable age information
for each individual cluster.  As discussed in Section~\ref{sec:hst_esc}, this 
is best done by including $U$-band imaging, which is costly to obtain in terms 
of observing time.  So far, MFs have only been constrained for a few, 
well-studied systems.  In the Antennae, \cite{zf99} found a power-law 
shape with exponent $\alpha\approx2$ over the mass range 
$10^4\,\msun$ to $10^6\,\msun$, similar to the MF of young LMC clusters. 
\cite{bik03} find $\alpha=2.1\pm0.3$ over the range $10^3\,\msun$ to 
$10^5\,\msun$ for M51, and \cite{deg03a} find $\alpha=2.04\pm0.23$ and
$\alpha=1.96\pm0.15$ in NGC~3310 and NGC~6745.

  The many studies which have found similar power-law \emph{luminosity} 
functions are of course consistent with these results, but should not be 
taken as proof that the MF is as universal as the LF. Conversely, any 
differences in the LFs observed in different systems would not necessarily 
imply that the MFs are different. There are some hints that slight LF 
variations may be present: \cite{elm01} find LF slopes of $\alpha=1.58\pm0.12$
and $\alpha=1.85\pm0.05$ in NGC~2207 and IC~2163, while \cite{lar02}
and \cite{dk02} find somewhat steeper slopes ($\alpha=2.0-2.5$) in several
nearby spiral galaxies. While \cite{whit99} find $\alpha=2.12\pm0.04$
for the full sample of Antennae clusters, there is some evidence for
a steepening at brighter magnitudes with $\alpha=2.6\pm0.2$ brighter than
$M_V=-10.4$. However, measurements of LF slopes are subject to many
uncertainties, as completeness and contamination effects can be difficult 
to fully control, and it is not presently clear how significant these 
differences are.  More data is needed.

  Another important question is how the MF evolves over time. While 
the evidence available so far indicates that the MF in most young cluster 
systems 
is well approximated by a uniform power-law with slope $\alpha\approx2$ down 
to the detection limit, old GC systems show a quite different behavior. Here, 
the luminosity function is well fit by a roughly log-normal distribution with 
a peak at $M_V\sim-7.3$ (about $10^5\,\msun$ for an age of 10--15 Gyr) and 
dispersion $\sim1.2$ mag (\cite[e.g.\ Harris \& van den Bergh 1981]{hv81}). 
Thus, \emph{old} globular clusters appear to have a characteristic mass
of about $\sim10^5\,\msun$, while there is no characteristic mass
for \emph{young} clusters. This difference might seem to imply fundamentally
different formation mechanisms. However, model calculations for the Milky Way
GC system by \cite{fz01} indicate that this difference can
be accounted for by dynamical evolution of the cluster system,
which makes the low-mass clusters disrupt more quickly and thereby
causes an initial power-law mass distribution to eventually approach
the bell-shaped MF seen in old GC systems.  Simulations by \cite{ves03} 
and \cite{vz03} suggest that an initial power-law MF can evolve towards 
a bell-shaped distribution also in ellipticals. 

  It is puzzling, however, that the ``faint fuzzy'' star clusters in 
NGC~1023 do \emph{not} show a turn-over in the MF at $\sim10^5\,\msun$  
even though they appear as old as the normal GCs in NGC~1023 (which do 
show the usual turn-over). Instead, there is a steady increase in the 
number of faint fuzzies at least down to the detection limit at $M_V\sim-6$ 
(\cite[Larsen \& Brodie 2000]{lb00}). It appears counter-intuitive that
these diffuse objects should be more stable against disruption than
compact GCs, although it may be significant that the faint fuzzies seem
to be on roughly circular orbits in the disk of NGC~1023
(Burkert et al., in preparation).

  Deriving a MF for an intermediate-age sample of clusters would
provide an important observational constraint on models
for dynamical evolution.  In the $\sim1$ Gyr fossil starburst M82B, 
the analysis by \cite{deg03b} indicates a turn-over in the MF 
at a mass of about $10^5\,\msun$, making the MF rather similar
to that of old globular clusters.  This would suggest that the
erosion of the MF is already well advanced at an age of $\sim10^9$ years 
in this system.
On the other hand, \cite{gou03} 
find no turn-over in the mass distribution of 3--Gyr old clusters in 
the merger remnant NGC~1316 down to their completeness limit at 
$M_B\approx-6$, or about 1 mag below the mass corresponding to the 
turn-over observed in old GC systems (accounting for evolutionary fading 
from 3 Gyr to 10 Gyrs). Thus, while it appears plausible that
the MF observed in old GC systems may indeed have evolved from an
initial power-law distribution as seen in young cluster systems,
more observational constrains would be highly desirable.  Here, 
observations with ACS or WFC3 would play a crucial role, since
high spatial resolution is required to detect the faintest clusters
and separate them from stars and background galaxies.

\subsection{Size-of-sample effects}
\label{sec:size_sample}

\begin{figure}
\centerline{
  \epsfxsize=14cm
  \epsfbox{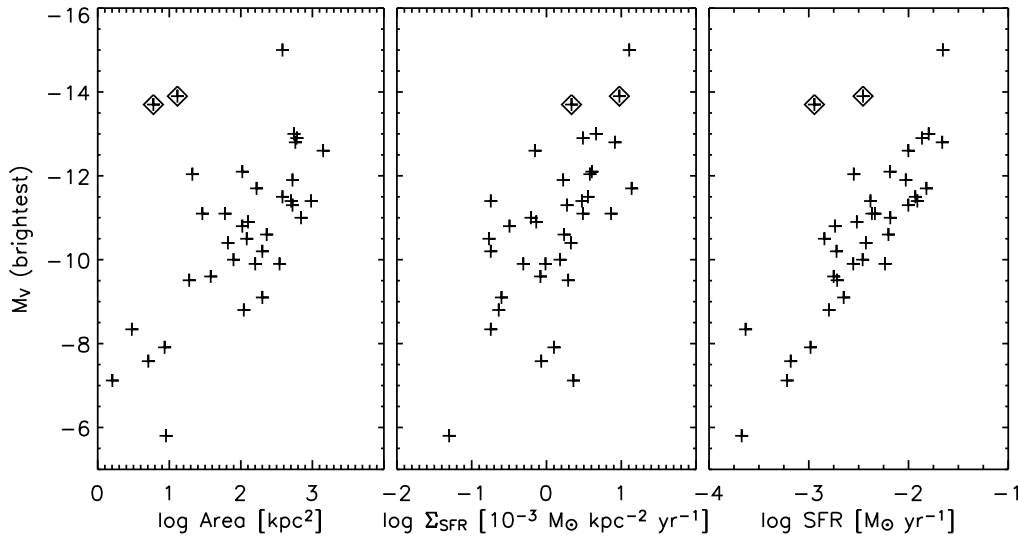}
}
\caption{\label{fig:lmax}Magnitude of brightest cluster versus total
  surveyed area (left), area-normalized star formation rate (center), and 
  total star formation rate (right). NGC~1569 and NGC~1705 are marked
  with diamonds.}
\end{figure}

  Because power-laws have no characteristic scale, it is hard to make 
a meaningful division between low-mass open clusters and higher-mass 
``super'' clusters in cluster systems with a power-law MF.
The lack of a characteristic mass suggests that there is
no fundamental difference between the physical processes behind
formation of clusters of various masses. Nevertheless, there
are evidently differences in the numbers of YMCs
(according to anyone's preferred definition) from one galaxy to
another. But how significant are these differences? And in
particular, is there an upper limit to the mass of a star cluster
that can form in a given galaxy?

  Size-of-sample effects may play
an important role in explaining the apparent differences between
cluster systems in different galaxies. As demonstrated by
\cite{whit03}, \cite{bil02} and \cite{lar02}, there is a strong 
correlation between the luminosity of the brightest cluster in
a galaxy and the total number of young clusters down to some magnitude
limit.  Moreover, this relation has 
the same form as one would expect if the luminosity function is 
a power-law where the maximum luminosity is simply dictated by 
sampling statistics (\cite[Fig.\ 10 in Whitmore 2003]{whit03}). Monte-Carlo
simulations indicate that the \emph{scatter} around the expected relation
(about 1 mag) is also consistent with sampling statistics 
(\cite[Larsen 2002]{lar02}).  In other words, current data are consistent 
with a universal, power-law luminosity function for young clusters in 
most galaxies, with the brightest clusters simply forming the tail of a
continuous distribution.  A similar analysis has yet to be carried out 
for the \emph{mass} distributions of star clusters in a significant 
sample of galaxies.

\begin{figure}
\centerline{
  \epsfxsize=11cm
  \epsfbox{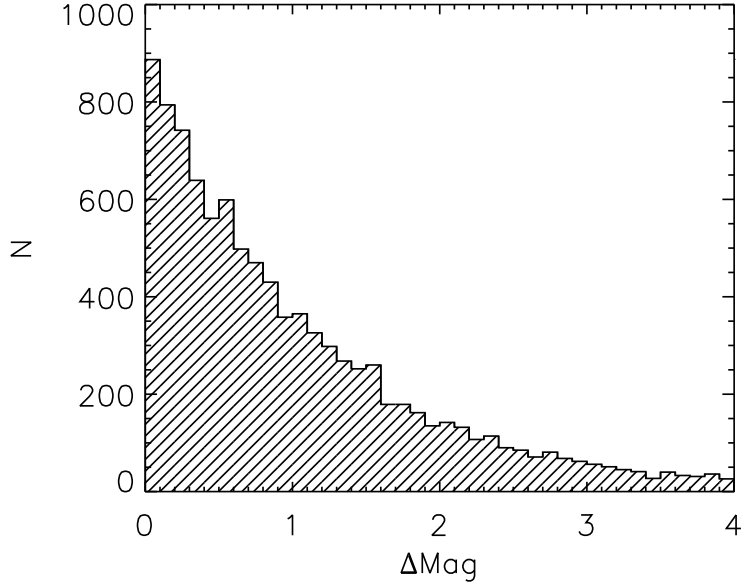}
}
\caption{\label{fig:dmag}Simulated distribution of the difference 
$\Delta$Mag, defined as the magnitude difference between two random clusters 
drawn from a power-law luminosity distribution, for 10000 experiments. 
This is equivalent to simulating the distribution of magnitude differences 
between the brightest and second-brightest cluster drawn from a power-law 
luminosity distribution in a sample of galaxies.  The
median $\Delta$Mag is 0.76 mag, while 25\%, 75\% and 90\% of the
experiments have $\Delta$Mag less than 0.31 mag, 1.50 mag and 2.51 mag.}
\end{figure}

  While the comparison of maximum cluster luminosity versus total 
cluster population is suggestive, it has its difficulties. Estimating
the total cluster population to some magnitude limit is subject to
uncertainties, due to the different completeness limits and detection
criteria applied in various studies.  \cite{lar02} attempted to circumvent 
this problem by assuming that the total cluster population scales with 
the galaxy SFR, but here the difficulty is to determine the proper 
normalization and account for the possibility that the scaling may
not be linear 
(\cite[Meurer et al.\ 1995; Larsen \& Richtler 2000]{meu95,lr00}).  

  Fig.~\ref{fig:lmax} shows the magnitude of the brightest cluster
$M_V^{\rm br}$ in the sample of galaxies studied by \cite{lar02}
versus total surveyed area $A$ (left panel), area-normalized star formation
$\Sigma_{\rm SFR}$ (center) and total star formation rate, 
SFR = $A\times\Sigma_{\rm SFR}$ (right). All three panels show some
degree of correlation, emphasizing the difficulty of disentangling
size-of-sample effects from physical effects. The most obvious
interpretation of the correlation between $M_V^{\rm br}$ and $A$ is
a purely \emph{statistical} one, i.e.\ larger galaxies have more clusters
on average, and therefore $M_V^{\rm br}$ becomes brighter, by
the size-of-sample effect. The $\Sigma_{\rm SFR}$ vs.\ $M_V^{\rm br}$
relation, on the other hand, is suggestive of a \emph{physical}
explanation: $\Sigma_{\rm SFR}$ correlates with the gas density, for
example (\cite[Kennicutt 1998]{ken98}), and the higher gas densities
and -pressures in galaxies with high $\Sigma_{\rm SFR}$ might provide 
conditions which are conducive for YMC formation 
(\cite[Elmegreen \& Efremov 1997]{ee97}). An important clue may lie
in the fact that the third plot, SFR vs.\ $M_V^{\rm br}$, shows the 
tightest correlation of all. This suggests that global galaxy properties
(SFR) are more important than local ones ($\Sigma_{\rm SFR}$) for a
galaxy's ability to form massive clusters. 

  An alternative metric is to simply compare the luminosities 
of the two brightest clusters in galaxies. If the luminosity function is
an untruncated power-law, this magnitude difference ($\Delta$Mag) should 
behave in a predictable way which can then be compared with observations. 
From an observational point of view, this approach has a number
of attractive features: only the two brightest clusters in a galaxy need 
be detected, distance uncertainties are irrelevant, and 
heterogeneous data (e.g.\ use of different bandpasses in different
galaxies) do not constitute a problem.  An obvious disadvantage is that 
this metric does not ``catch'' cases like NGC~1569, where the magnitude 
difference between the two brightest clusters will not reveal the gap 
down to number 3. Also, a large sample of galaxies is needed to get 
meaningful results.

\begin{figure}
\centerline{
  \epsfxsize=11cm
  \epsfbox{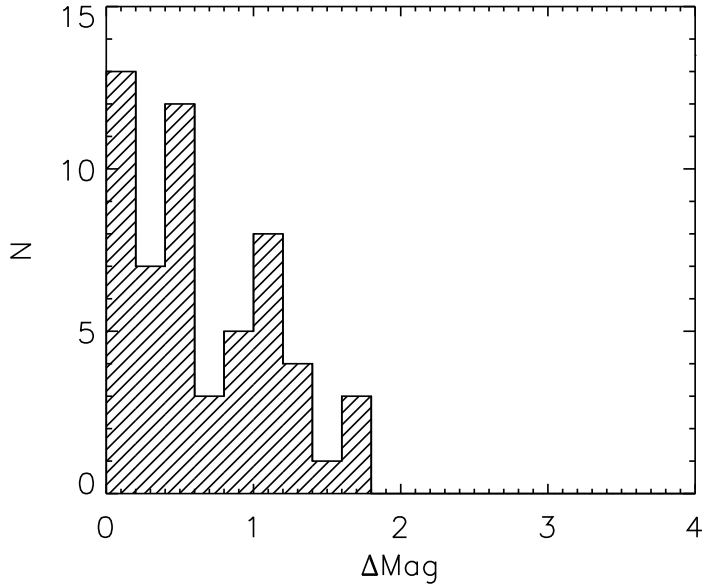}
}
\caption{\label{fig:dmobs}Observed distribution of $\Delta$Mag values
for 51 galaxies.
}
\end{figure}

  Figure~\ref{fig:dmag} shows a simulation of the distribution of 
$\Delta$Mag, obtained in a series of 10000 experiments where cluster 
populations 
were drawn at random from a power-law distribution with exponent 
$\alpha=2$.  Fig.~\ref{fig:dmobs} shows the distribution of observed
$\Delta$Mag values for a sample of 57 galaxies, consisting of the spirals 
and dwarf galaxies surveyed by \cite{lr99} and \cite{bil02} and
data from the sources 
listed in Tables~\ref{tab:nmsb}--\ref{tab:sp} (galaxies marked with
asterisks) where photometry is given for 
individual clusters. The observed $\Delta$Mag distribution in 
Fig.~\ref{fig:dmobs} is qualitatively similar to that in Fig.~\ref{fig:dmag}, 
peaked towards $\Delta$Mag=0 and with a tail extending up to larger
$\Delta$Mag values. Quantitatively, a Kolmogov-Smirnov test returns 
only a 3\% probability that the distribution in Fig.~\ref{fig:dmobs}
is drawn from that in Fig.~\ref{fig:dmag}, but the agreement
improves greatly for a somewhat steeper power-law slope: For $\alpha=2.2$
and $\alpha=2.4$, the K-S test returns a probability of 30\%
and 77\% that the observed $\Delta$Mag distribution is consistent
with the equivalent of Fig.~\ref{fig:dmag}. Since this comparison
is only sensitive to the very upper end of the LF, it may indicate
that the slope at the bright end of the LF is typically somewhat steeper than
$\alpha=2$.

  One remaining issue is the apparently discontinuous luminosity function 
in some dwarf galaxies, where the brightest clusters are much \emph{too 
luminous} for the total number of clusters. It has been suggested that
these cases may represent a mode of star formation which is distinct from 
that operating in larger galaxies, possibly caused by transient high-pressure
disturbances (\cite[Elmegreen 2004]{bge04}).  Again, it may be worth asking 
how significant these exceptions are.  The median $\Delta$Mag from 
Fig.~\ref{fig:dmag} is 0.76 mag, but in 25\% of the cases there is a 
difference of $\Delta$ Mag = 1.50 mag or more, and a $\Delta$ Mag $>$ 2.5 
mag is found in 10\% of the cases. Thus, it is not entirely unlikely to 
find a gap of 2 mag from the brightest to the second-brightest cluster. 

  One might argue that it does seem unlikely to form \emph{two} or 
more very massive clusters by chance.  However, the luminosity- (or mass) 
function is a statistical tool which may not apply within small regions, 
but only when averaged over an entire galaxy.  
A dwarf galaxy like NGC~1569 may be considered as essentially 
a single starburst region, whereas a larger galaxy contains a multitude
of starforming regions of various sizes.  The experiment on which 
Fig.~\ref{fig:dmag} is based does not apply within a region that has been 
selected \emph{a priori} to contain a massive cluster. For example, if
one selects a small subregion around the brightest cluster in a large
galaxy, it would be very unlikely to find the second-brightest cluster
in the galaxy within that subregion by chance.  
  A related question is whether the formation of massive clusters is
correlated -- is there an increased probability of finding one massive
cluster forming next to another one? 
The answer appears to be affirmative. In addition to the case of
NGC~1569-A, there are several examples of binary
clusters with roughly equal masses in the Large Magellanic Cloud 
(\cite[Dieball et al.\ 2002]{di02}). Closer to home, the famous 
``double cluster'' $h$ and $\chi$ Per is another example. The
double cluster, incidentally, is also among the most luminous open clusters
known in the Milky Way.

%
%

\subsection{Cluster sizes}

  Another point hinting at a universal cluster formation mechanism
is the observation that most star clusters seem to have about the same
size. The initial WFPC data for young clusters in the Antennae
indicated a mean half-light radius of about 18 pc (\cite[Whitmore \&
Schweizer 1995]{ws95}), rather large compared to the $\sim$3 pc typical 
of old globular clusters. This was used as an argument against the
idea that the young clusters in the Antennae are young globular clusters
(\cite[van den Bergh 1995]{van95}), but with WFPC2 data the mean size was
revised to $4\pm1$ pc (\cite[Whitmore et al.\ 1999]{whit99}). Similar results 
have since been found for star clusters in many other galaxies (see examples 
in preceding sections). The discrepancy between size measurements of
Antennae clusters carried out on WFPC and WFPC2 images illustrates the 
importance of having sufficient spatial resolution, however.

  Naively, one might expect clusters to form with a constant
density rather than a constant size, but intriguingly, there is little 
to suggest any significant size-mass
correlation for star clusters.  This has now been demonstrated both in young 
cluster systems in starburst regions 
(\cite[Zepf et al.\ 1999; Carlson \& Holtzman 2001]{zep99,ch01}), in 
spiral galaxies (\cite[Larsen 2004]{lar04}), for old globular clusters
in the Milky Way (\cite[van den Bergh et al.\ 1991]{van91}) and
elsewhere (\cite[Kundu \& Whitmore 2001]{kw01}), and even in
nuclear star clusters (\cite[B{\"o}ker et al.\ 2004]{bok04}). In the absence 
of any such relation, it is clear that the most massive clusters will also 
tend to have very high densities, and explaining these high densities is
therefore just a special case of the more general problem why
star clusters form with a nearly constant size,
rather than with a constant density. \cite{az01} proposed that
the lack of a size-mass relation may be related to a higher star formation
efficiency in high-mass clusters, causing them to expand less than 
lower-mass clusters after the residual gas is blown away, but more work
is required to obtain a deeper understanding of this issue.

  The lack of a size-mass relation is all the more puzzling because 
there \emph{are} real variations in the sizes of star clusters. The 
half-light radii of old globular clusters correlate with galactocentric 
distance (\cite[van den Bergh et al.\ 1991; McLaughlin 2000]{van91,mc00}), 
and some lenticular galaxies have ``faint fuzzy'' star clusters with much
larger sizes than normal open and globular clusters (\cite[Larsen \& 
Brodie 2000]{lb00}).  This issue clearly merits further investigation.

\subsection{Efficiency of cluster formation and disruption}
\label{sec:eff}

  Many of the studies cited in Section~\ref{sec:env} have found that a 
high fraction of the luminosity in starburst regions is emitted by clusters
or compact sources, although a direct intercomparison is difficult because of 
the different bandpasses used (UV through IR) and different 
detection limits. \cite{meu95} warn that some of their compact sources may
not be individual star clusters, and similar caution applies in other
systems, especially as observations are pushed to greater distances
where the nature of cluster candidates cannot be 
verified because of insufficient resolution.  Nevertheless, there are strong 
indications that most stars tend to form in groups or clusters, although only a 
small fraction of all stars eventually end up in \emph{bound} clusters.  
In the Milky Way, \cite{ll03} estimate that the vast majority (70\%-90\%) of 
star formation in nearby molecular clouds takes place in embedded clusters, 
while only 4\%--7\% of these embedded clusters survive to become bound 
clusters with 
ages similar to that of the Pleiades ($\sim10^8$ years).  Similar results
have been found in other galaxies: In NGC~5253, \cite{tre01} found their 
data to be consistent with a scenario where all stars are initially born in 
clusters, of which most disperse on a short time scale ($\sim10$ Myrs). In 
the Antennae, \cite{fall04} estimates that at least 20\%, and possibly all 
star formation takes place in clusters, although he also concludes that most 
are unbound and short-lived. This also seems consistent with the finding by 
\cite{kc88} (Sec~\ref{sec:early}) that only a small fraction of giant 
H{\sc ii} regions in late-type galaxies form massive, bound clusters, while 
the rest are forming unbound associations. 

  It is unclear how these findings relate to the apparent variations in 
specific cluster luminosity $T_L$ with host galaxy star formation rate 
(\cite[Larsen \& Richtler 2000]{lr00}). These authors found that the 
fraction of $U$-band light from star clusters relative to their host 
galaxies increases with the area-normalized SFR, from 0.1\%--1\% for most 
normal spiral galaxies up to the very large fractions (20\%--50\%) found 
in starbursts. If nearly all stars initially form in clusters, $T_L$ may 
reflect a survival- rather than a formation efficiency. However, other 
factors such as dust extinction and the details of the star formation 
history of the host galaxy could also affect $T_L$. Within an on-going, 
strong starburst, a large fraction of the light comes from clusters, 
leading to a high $T_L$. After the burst, fading alone would not affect 
$T_L$ (since field stars and clusters will fade by the same amount) but 
cluster disruption would cause $T_L$ to gradually decrease by an amount
that depends on the disruption timescale (\cite[Boutloukos \& Lamers
2003]{bl03}).

  Cluster disruption occurs on several timescales
(\cite[Whitmore 2004]{whit04}). Initially, the star formation efficiency 
(SFE) within the parent molecular cloud is a critical factor in determining
whether or not a given embedded cluster remains bound.  Unless the SFE exceeds 
30\%--50\% (\cite[Hills 1980; Boily \& Kroupa 2003]{hil80,bk03}), the cluster 
is likely to become unbound once the gas is expelled. Even when the SFE
is high, a large fraction of the stars may be unbound
and eventually disperse away (\cite[Kroupa et al.\ 2001]{kro01}). The
high ``infant mortality'' for clusters may represent a combination of 
rapid disruption of clusters which are unbound altogether and clusters which 
retain a bound core containing only a small fraction of the initial mass, 
but interestingly, this effect appears to be largely independent of cluster 
mass (\cite[Fall 2004]{fall04}).  On longer timescales, clusters continue
to disrupt due to two-body relaxation, tidal shocks and other effects
(\cite[Boutloukos \& Lamers 2003]{bl03}) which may be instrumental in 
shaping the globular cluster mass function (Section~\ref{sec:lfmf}). 

\subsection{Dynamical Masses and the Distribution of Stellar Masses}

  A somewhat controverial question, which is related to the disruption 
timescales, concerns the universality of the stellar mass function
(SMF) in YMCs. Note
that the term ``IMF'' (initial mass function) is deliberately avoided
here, since the present-day mass function in a star cluster may differ
from the initial one. If the SMF is biased towards high-mass stars,
compared e.g.\ to a \cite{kro02}-type function,
the clusters might be rapidly disrupted (\cite[Goodwin 1997]{goo97}).
Direct observations of individual stars in YMCs are usually beyond reach 
even with HST, especially at the low-mass end of the SMF.
However, by measuring structural parameters on HST images and using 
ground-based high-dispersion spectra to estimate the line-of-sight velocity 
dispersions of the cluster stars, dynamical mass estimates can be obtained
by simple application of the virial theorem. If the cluster ages are
known, the masses thus derived can 
be compared with SSP model calculations for various SMFs. An increasing 
amount of such data is now becoming available, but the results remain
ambiguous.  Based on observations by 
\cite[Ho \& Filippenko (1996a,1996b)]{hf96a,hf96b}, \cite{st98} found some 
evidence for differences in the SMF slopes of NGC1569-A (SMF at least as 
steep as a Salpeter law down to 0.1$\msun$) and NGC1705-1 (SMF may be shallower 
than Salpeter or truncated between $1\msun$ and $3\msun$), and \cite{sg01} 
also concluded that M82-F has a top-heavy 
SMF. However, \cite{gg03} found M/L ratios consistent with ``normal''
SMFs for three clusters in NGC~1569, \cite{mar04} reported
``excellent'' agreement between the dynamical mass of the W3 object in 
NGC~7252 and SSP model predictions, and \cite{lr04} and \cite{lbh04} found 
standard Kroupa (2002)-like SMFs for 7 YMCs in a sample of dwarfs and 
spiral galaxies.  Other authors have found a mixture of normal and 
top-heavy SMFs (\cite[Mengel et al.\ 2002; McCrady et al.  2003]{men02,mgg03}).
At present, it is not clear to what extent these differences are real,
or could be a result of different measurement techniques, crowding and
resolution effects, as well as the inherent uncertainties in the analysis
(e.g.\ assumption of virial equilibrium, effects of mass segregation, different
macroturbulent velocities in the atmospheres of the cluster supergiants 
and the template stars). However, this aspect of YMC research would have
been virtually impossible without the symbiosis between HST 
and high-dispersion spectrographs large ground-based telescopes.

\section{Summary and outlook}


  Over the past decade, research in extragalactic young
star clusters has evolved into a mature field. This is in no small part
due to the complement of instruments available on HST, although the
impact of parallel developments in ground-based instrumentation should not 
be underestimated. 

  One major advance has been to establish the ubiquity of young massive,
globular-like clusters in a wide variety of star-forming environments.
These objects can have sizes and masses which make them virtually identical
to the old globular clusters observed in the Milky Way and indeed
around all major galaxies. 

  With few exceptions, the luminosity functions of 
young cluster systems are well approximated by a power-law with exponent
$\sim2$. It appears that random sampling from such a luminosity function 
can account, to a large extent, for differences in the numbers of
YMCs and in the luminosity of the brightest star clusters observed in 
different galaxies.  So far, no case has been found in which the luminosity
of the brightest cluster is limited by anything other than sampling
statistics.  In other words, \emph{YMCs are present whenever
clusters form in large numbers}. Studies of the \emph{mass} functions of young
star clusters are more difficult, but the few studies that have been
made seem to indicate that the mass functions are also well approximated
by power-laws.  

  It remains unclear to what extent dwarf galaxies like NGC~1569 and 
NGC~1705, with only a few very bright clusters, pose a problem for the 
idea of a universal cluster luminosity (or mass) function and a
universal cluster formation mechanism.  \cite{bil02} noted that 
massive star clusters are rare even in actively star forming dwarfs,
but when they do form they are accompanied by a high level
of star formation activity.  Billett et al.\ suggested that 
large-scale flows and gravitational instabilities in the absence of
shear may favor the formation of massive clusters in dwarf galaxies.

  By focusing on the most extreme starburst environments, one naturally finds 
the most extreme cluster populations. However, a complete picture can only 
be formed by examining the whole range of environments, from very quiescent, 
over normal star-forming galaxies, to starbursts, and good progress is being 
made towards this goal.  An example of the first extreme is IC~1613, whose 
extremely low (but non-vanishing) star formation rate has produced 
only a very feeble cluster system. Even among ``normal'' spiral
galaxies there are substantial variations in the SFR, and this translates 
directly to corresponding differences in the richness of the cluster 
systems. At the other extreme are starbursts like the Antennae and
M82, with their exceedingly rich young cluster systems.

While special conditions may lead to the formation of a few massive clusters 
in some dwarfs, YMC formation apparently does not \emph{require} special 
triggering mechanisms. This is supported by the study of young clusters
in the Antennae by \cite{zha01}, who found no correlation between
strong gradients in the velocity field and the formation sites of
star clusters, although it was noted that \emph{some} clusters might
have been triggered by cloud-cloud collisions.  By a slight extrapolation
of this argumentation, globular clusters could also have formed by 
normal star formation processes in the early Universe.



  A large fraction, possibly
the majority, of all stars form in star clusters. However, only a
small fraction of these stars eventually end up as members of
\emph{bound} star clusters, due to an initially rapid disruption of 
clusters on timescales of $10^7$ years which appears to be largely
independent of mass. Clusters which survive the initial phase of
rapid destruction continue to dissolve on longer timescales, and
this process may cause an initial power-law mass function to
evolve towards the approximately log-normal MF observed in old
globular cluster systems.

  Further progress is likely to come from multi-wavelength campaigns,
which will allow detailed analyses of the age- and mass distributions of
cluster systems.  Is there a universal cluster mass function, as hinted at 
by the many studies which have found similar power-law luminosity functions? 
Is the ``infant mortality'' rate
everywhere the same?  Can we see the signatures of
dynamical evolution?  A good place to look for such signatures might
be in galaxies with rich cluster systems formed over an extended
period of time.  Also, what are the properties of young clusters in
early-type (Sa/Sb) spirals and other environments (e.\ g.\
low surface brightness galaxies) which remain poorly studied? 


  Over its lifetime HST has been upgraded several times, each time
essentially leaving us with a new, much more capable observatory. 
The Wide Field Planetary Camera 2 has now been in operation for
more than 10 years, and has produced many spectacular results.  We have 
only just started to see the potential of the \emph{Advanced Camera for 
Surveys}, and the \emph{Wide Field Camera 3} with its panchromatic 
wavelength coverage from the near-ultraviolet to the infrared has the
potential to once more boost the power of Hubble by a significant
factor.

\end{document}